\documentclass[aps,pra,twocolumn,superscriptaddress]{revtex4-1}

\usepackage{graphicx}
\usepackage{hyperref}
\usepackage{amsmath}
\usepackage{amssymb}
\usepackage{epstopdf}
\usepackage[usenames]{color}

\begin{document}

\title{Mid-infrared single-photon 3D imaging}

\author{Jianan Fang}
\affiliation{State Key Laboratory of Precision Spectroscopy, East China Normal University, Shanghai 200062, China}

\author{Kun Huang}
\email{khuang@lps.ecnu.edu.cn}
\affiliation{State Key Laboratory of Precision Spectroscopy, East China Normal University, Shanghai 200062, China}
\affiliation{Chongqing Key Laboratory of Precision Optics, Chongqing Institute of East China Normal University, Chongqing 401121, China}
\affiliation{Collaborative Innovation Center of Extreme Optics, Shanxi University, Taiyuan, Shanxi 030006, China}

\author{E Wu}
\affiliation{State Key Laboratory of Precision Spectroscopy, East China Normal University, Shanghai 200062, China}
\affiliation{Chongqing Key Laboratory of Precision Optics, Chongqing Institute of East China Normal University, Chongqing 401121, China}

\author{Ming Yan}
\affiliation{State Key Laboratory of Precision Spectroscopy, East China Normal University, Shanghai 200062, China}
\affiliation{Chongqing Key Laboratory of Precision Optics, Chongqing Institute of East China Normal University, Chongqing 401121, China}

\author{Heping Zeng}
\email{hpzeng@phy.ecnu.edu.cn}
\affiliation{State Key Laboratory of Precision Spectroscopy, East China Normal University, Shanghai 200062, China}
\affiliation{Chongqing Key Laboratory of Precision Optics, Chongqing Institute of East China Normal University, Chongqing 401121, China}
\affiliation{Chongqing Institute for Brain and Intelligence, Guangyang Bay Laboratory, Chongqing, 400064, China}
\affiliation{Shanghai Research Center for Quantum Sciences, Shanghai 201315, China}
\affiliation{Jinan Institute of Quantum Technology, Jinan, Shandong 250101, China}

\begin{abstract}
Active mid-infrared (MIR) imagers capable of retrieving three-dimensional (3D) structure and reflectivity information are highly attractive in a wide range of biomedical and industrial applications. However, the infrared 3D imaging at low-light levels is still challenging due to the deficiency of sensitive and fast MIR sensors. Here we propose and implement a MIR time-of-flight imaging system that operates at single-photon sensitivity and femtosecond timing resolution. Specifically, back-scattered infrared photons from a scene are optically gated by delay-controlled ultrashort pump pulses through nonlinear frequency upconversion. The upconverted images with time stamps are then recorded by a silicon camera to facilitate the 3D reconstruction with high lateral and depth resolutions. Moreover, an effective numerical denoiser based on spatiotemporal correlation allows us to reveal the object profile and reflectivity under photon-starving conditions with a detected flux below 0.05 photons/pixel/second. The presented MIR 3D imager features with high detection sensitivity, precise timing resolution, and wide-field operation, which may open new possibilities in life and material sciences.
\end{abstract}

\maketitle

\vspace{8pt}
\noindent{\fontfamily{phv}\selectfont 
\textbf{Introduction}}
\vspace{4pt}
\newline
Laser-based three-dimensional (3D) imaging plays an essential role in widespread applications, ranging from biomedical diagnosis, remote sensing to target recognition and identification \cite{Drouin2012Chapter, Li2020Optica, Sun2013Science, Kirmani2014Science}. Recently, the 3D imaging technology has been greatly fueled by the emergence of improved hardwares, novel configurations and effective algorithms  \cite{Drouin2012Chapter}, which significantly improves the key performances for laser raging \cite{Jiang2020NP}, surface profiling \cite{Kim2022NC} and volumetric tomography \cite{Siddiqui2018NP}. In particular, low-light capabilities in the sparse photon regime and ultrahigh temporal resolutions at the picosecond level can readily be achieved by using computation-assisted advanced imagers, for instance based on single-photon avalanche diode (SPAD) arrays \cite{Gariepy2015NC,Shin2018NC,Morimoto2020Optica}, superconducting nanowire sensors \cite{Zhao2017NP,Kong2020OL}, streak cameras \cite{Velten2012NC,Gao2014Nature}, and intensified charge-coupled devices (ICCDs) \cite{Morris2014NC,Faccio2020NRP}. However, these state-of-the-art 3D imaging performances are so far limited in the visible or near-infrared regions. Nowadays, there is a significant impulse to extend the operation wavelength into the mid-infrared (MIR) region, pertaining to unique features of the chemical selectivity with inherent rovibrational contrast and the enhanced penetration depth with reduced photon scattering \cite{Vodopyanov2020Book}. Therefore, there is an urgent need for developing high-dimensional MIR imaging technologies with high sensitivity and high resolution, with an aim to promote broader applications related to chemical, medical, and biological fields \cite{Guo2002OE,Martin2013NM}.

Typically, MIR tomographic imaging can be performed by optical coherent tomography (OCT), which has been recognized as an indispensable tool to identify morphological structures and subsurface defects \cite{Colley2007RSI,Su2014OE}. However, the current sensitivity of the OCT system struggles to examine weakly reflective interfaces of thick diffusive materials or highly absorptive constituents in aqueous biological samples \cite{Israelsen2019LSA, Zorin2018OE}, thus devaluating the low-scattering advantage at long operation wavelengths. One prerequisite for approaching rapid and sensitive OCT measurements is to realize the high-speed and low-noise detection of infrared radiation, which remains a challenge for existing MIR detectors \cite{Razeghi2014RPP}. Indeed, the sensitivity of conventional MIR detectors based on narrow-bandgap semiconductors is severally limited by the intrinsic dark current, which usually requires cryogenic working conditions at the expense of additional cost and complexity \cite{Wang2019Small}. Notably, emerging low-dimensional materials constitute promising platforms to implement room-temperature MIR imagers, albeit that the detection sensitivity is far from the single-photon level \cite{Long2019AFM}. Currently, the large-area deposition with high-density pixels is impeded by the complicated and expensive fabrication technology \cite{Wu2021NR}. In addition, the response time for the pixelated device is usually longer than microseconds, which is restricted by the lifetime of the photo-carriers and the bandwidth of readout circuits \cite{Wang2019Small}. Therefore, it is still a long-standing goal to realize direct MIR imagers with high detection sensitivity, massively parallel elements, and fast temporal resolution.

Alternatively, an indirect approach has been developed to address the aforementioned limitations of arrayed MIR detectors. In this approach, an intermediate conversion step is used to transfer the spatio-temporal information encoded in the MIR spectrum into the visible replica, where high-performance silicon detectors can be used to leverage the desirable features of high detection efficiency, low dark noise, and high-density pixels. So far, there have been various schemes proposed to realize the MIR 3D imaging. For instance, MIR photothermal imaging is implemented by using a visible laser to probe the infrared absorption-induced thermal lensing effect in the sample \cite{Zhang2016SA, Tamamitsu2020Optica}. This technique has proven to be useful to label-free 3D chemical imaging of live cells and organisms \cite{BaiSA2021}, yet it is only applicable to microscopic translucent samples, and requires high-photon-flux infrared illumination at the mW level \cite{Zhang2016SA, Tamamitsu2020Optica, BaiSA2021}. Another promising method is to use the optical nonlinearity of the detector material based on non-degenerate two-photon absorption \cite{Fishman2011NP}. The wide-field imaging method is featured with high-definition and high speed \cite{Potma2021Optica,Potma2021APLP}, albeit that the detection sensitivity is intrinsically limited by the low efficiencies of the third-order nonlinear process. The increase of the pump intensity would be detrimental for improving the signal-to-noise (SNR) ratio due to the more rapid augmentation of background noises from the harmonic pump absorption \cite{Fang2020PRAp}. Consequently, further reduction of the required MIR illumination flux calls for highly-sensitive MIR imaging techniques at the single-photon level, which is potentially beneficial to low-light-level scenarios, such as non-destructive testing in cultural heritage conversation, remote sensing at a long standoff distance, and phototoxicity-free observation of biomedical samples \cite{Altmann2018Science}.

In this context, upconversion imaging based on coherent parametric conversion provides an effective routine to realize MIR single-photon 3D imaging due to the intrinsically noise-free operation \cite{Barh2019AOP}. The combination of the nonlinear frequency conversion with OCT techniques leads to the realization of sensitive MIR volumetric imaging \cite{Israelsen2019LSA}. Furthermore, the stringent requirement of broadband MIR light sources is mitigated by using the wide-band entangled photon pairs, which allows to realize time-domain \cite{Paterova2018QST} and frequency-domain \cite{Vanselow2020Optica} OCT systems based on single-photon nonlinear interferometry. Despite tremendous efforts, the presented imaging modality based on the OCT architecture faces the challenges that the total acquisition time is rather long due to the lateral raster scanning fashion. In this regard, wide-field operation based on a silicon camera is developed in the framework of parametric upconversion imaging, which favors high-pixel density mapping with fast data acquisition and quick processing speed \cite{Junaid2019Optica, Zhang2022PR}. The snapshot capability allows one to capture the spatial information with an exposure time down to the microsecond level \cite{Huang2022NC}. Naturally, it is thus appealing to extend the MIR upconversion imaging technique to higher dimensions, which may pave a new way to addressing the long-sought-after goal for realizing the wide-field MIR 3D imaging at the single-photon level.

In this work, we have proposed and implemented a wide-field MIR 3D imaging system, which is featured with single-photon sensitivity, megapixel high definition, and femtosecond timing resolution. The underlying mechanism relies on the coincidence-pumping parametric upconversion imaging, where reflected infrared photons from a scene are optically gated by an ultrafast pulse before being captured by a silicon camera. The recorded slices with precise time stamps allow us to retrieve the structure and reflectivity information in a fast wide-field fashion. The range-gated operation strategy renders the MIR imaging system analogous to the ICCD camera, but offering a much shorter gate window with a femtosecond timing precision, far beyond the achievable electronic counterpart. Moreover, the nonlinear coincidence pumping involved in the frequency upconversion process facilitates to achieve a high conversion efficiency due to the intensive peak power of the pulsed pump, while significantly suppressing the background noises within an ultrashort time window. In addition, an effective computation algorithm is used to retrieve the photon-sparse information by leveraging the spatial and temporal correlations of a target. As a result, an ultra-sensitive depth and reflectivity imaging is manifested at the photon-starving regime with an extremely-low detected flux about 0.05 photons/pixel/second. It is anticipated that the MIR single-photon 3D imaging system would immediately stimulate subsequent low-light-level applications requiring high detection sensitivity and high temporal resolution.

\begin{figure*}[t!]
\centering
\includegraphics[width=0.9 \textwidth]{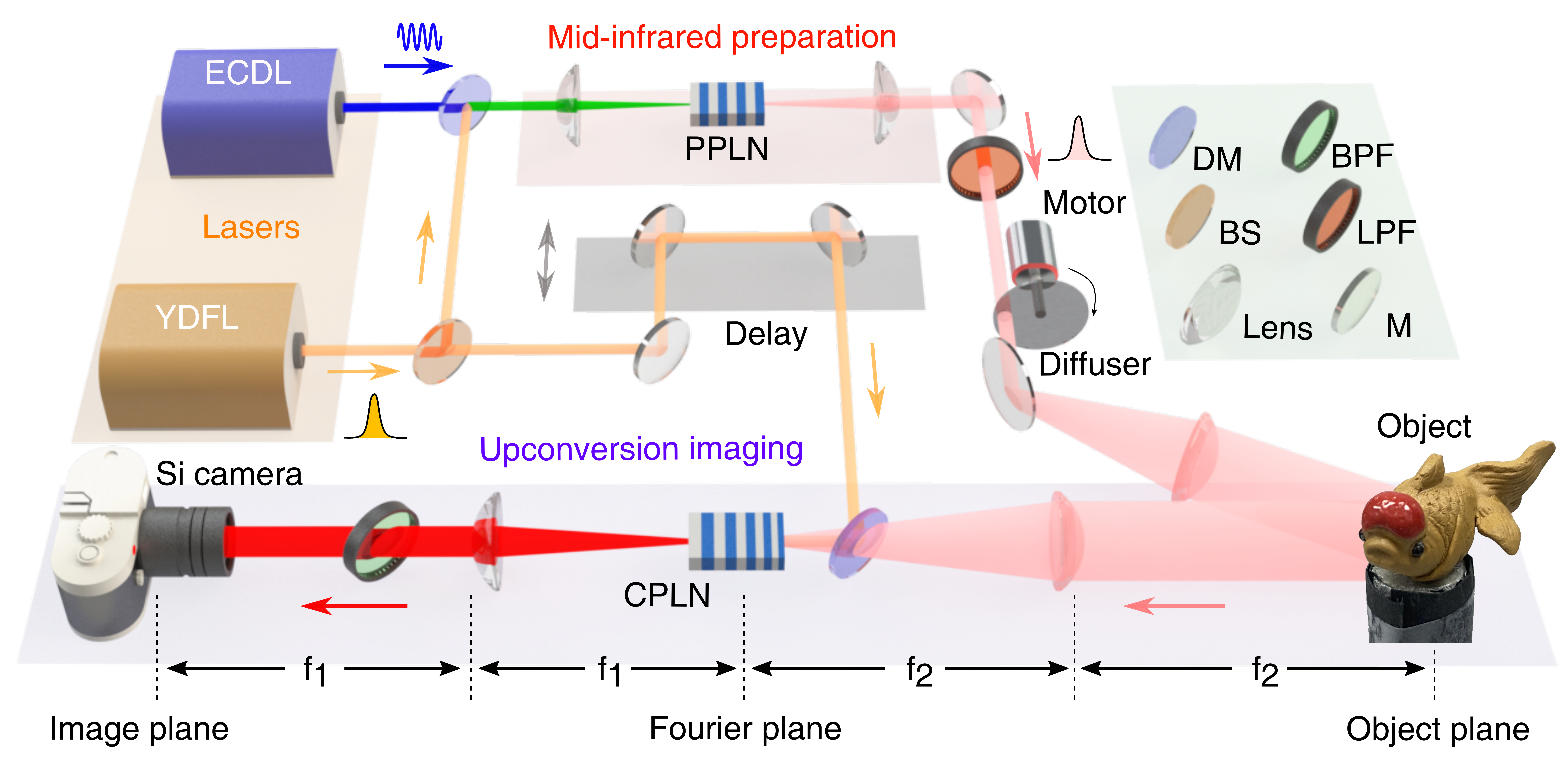}
\caption{\textbf{Experimental setup of the MIR single-photon upconversion 3D imaging.} The involved light sources originate from an ytterbium-doped fiber laser (YDFL) and an extended cavity diode laser (ECDL), which spectrally center at 1030 and 1550 nm, respectively. The YDFL delivers a train of mode-locked optical pulses, while the ECDL operates at the continuous-wave mode. The dual-color beams are then spatially combined by a dichroic mirror (DM) and injected into a periodically poled lithium niobate (PPLN) crystal to perform the difference-frequency generation for preparing the MIR pulsed illumination source. The signal beam passes through a spinning ground-glass diffuser before illuminating the object to smooth the speckles arising from the spatial coherence. Subsequently, the back-scattered infrared photons from the object are steered into a 4f upconversion imaging configuration, where a chirped-poling lithium niobate (CPLN) crystal is placed at the Fourier plane to realize the wide-field frequency upconversion. The upconverted image is captured by a silicon camera after a series of spectral filters. The time-resolved measurement for the tomographic imaging is realized by the nonlinear optical gating based on delay scanning of the synchronized pump pulse. The high slicing precision and high imaging sensitivity enable us to reconstruct the 3D profile and surface reflectivity of the targeted scene at the low-photon-flux regime.}
\label{fig1}
\end{figure*}

\vspace{8pt}
\noindent{\fontfamily{phv}\selectfont 
\textbf{Results}}
\vspace{4pt}
\newline
\textbf{Experimental setup.} Figure \ref{fig1} presents the experimental setup for the MIR single-photon upconversion 3D imaging. The involved time-of-flight modality is performed by incoherently illuminating a scene with ultrashort MIR pulses and recording the back-scattered light based on the frequency upconversion detection. The underlying core of the operation mechanism lies in the nonlinear optical gating based on coincidently pulsed pump, which facilitates time-resolved and sensitive MIR imaging. Specifically, the laser sources used in the experiment consist of an ytterbium-doped fiber laser (YDFL) and an extended cavity diode laser (ECDL). The YDFL is mode-locked at the repetition rate of 21.6 MHz to deliver ultrashort pulses at 1030 nm. The pulse duration after two cascaded fiber amplifiers is compressed to 270 fs by using a pair of dispersive gratings. A portion of the YDFL output is spatially combined with the continuous-wave ECDL at 1550 nm by a dichroic mirror. The mixed beams are injected into a periodically poled lithium niobate (PPLN) crystal via an achromatic lens, which allow us to prepare a MIR pulsed source at 3070 nm via difference-frequency generation (DFG). The generated MIR light is collimated by a calcium fluoride lens, and passes through a spinning diffuser to remove the speckle effect originated from the high spatial coherence of the laser source. After reflecting from the object surface, the diffused infrared photons are collected into a 4f upconversion imaging system, where a chirped-poling lithium niobate (CPLN) crystal is placed at the Fourier plane to perform sum-frequency generation (SFG). The CPLN is fabricated with a linearly-ramping poling period from 16 to 24 $\mu$m, which permits a wide-field upconversion imaging due to the self-adapted quasi-phase matching for various incident angles \cite{Huang2022NC}. After a spectral filtering, the upconverted image at 771 nm is recorded by an electron multiplying CCD (EMCCD). The associated time tag for each optical slice is determined by the temporal scanning delay of the ultrafast pump pulse. Consequently, an active imaging is implemented to retrieve the structure and reflectivity information of the 3D scenery. More details about the imaging system are given in Supplementary Note 1.

Notably, there are several distinct features in system designing to optimize the MIR 3D imaging performance. First, the coincident pulse pumping in the implemented MIR upconverter enables to realize the required time-resolved capability for the time-of-flight imaging, which is in marked contrast to the continuous-pumping configuration \cite{Dam2012NP}. The pulsed pumping favors to increase the conversion efficiency due to intensive peak power, and meanwhile to suppress the background noise with the narrow time window \cite{Huang2021PR, Donnell2019PR, Mrejen2020LPR}. In addition to improving the SNR ratio for MIR upconversion detection, the pump pulse serves as an ultrafast optical gate for the upconversion imager, thus allowing to identify incoming photons with different time delays \cite{Zhang2022PR, Rehain2020NC}. Second, the temporal resolution of the imaging system is determined by the peak width of the intensity cross-correlation trace between the pump and probe pulses, instead of being restricted to the response time of the detector itself. Here, the pulse duration of the pump pulse is optimized at the femtosecond regime, which significantly improves the depth resolution in comparison to previous demonstrations with picosecond pulses \cite{Huang2022NC, Zhang2022PR, Rehain2020NC}. On the other hand, the walk-off effect within the nonlinear crystal becomes more prominent for shorter pulses \cite{Donnell2019PR, Mrejen2020LPR}, which may reduce the conversion efficiency as a trade-off. As discussed in Supplementary Note 2, the pulse parameters used in our experiment ensure a negligible efficiency reduction below 3\%. Third, the wide-field upconversion is implemented to achieve a full imaging capability, which simplifies the data acquisition and reduces the acquisition time in comparison to the raster-scanning performance \cite{Rehain2020NC}. Moreover, the use of spinning diffuser allows us to implement an incoherent imaging based on the pseudo-thermal source \cite{Junaid2019Optica}, thus resulting in an enhanced the spatial resolution than that in the coherent imaging regime \cite{Zhang2022PR, Huang2022NC}. Fourth, the self-synchronization preparation for the dual-color light sources with disparate wavelengths eliminates the relative timing jitter between the signal and pump pulses. The gating functionality with a high precision and a high stability is particularly important to the low-light-level 3D imaging that usually takes a long exposure time to accumulate sufficient photons for high-quality image reconstruction \cite{Gariepy2015NC}.

\begin{figure}[b!]
\centering
\includegraphics[width=0.75\columnwidth]{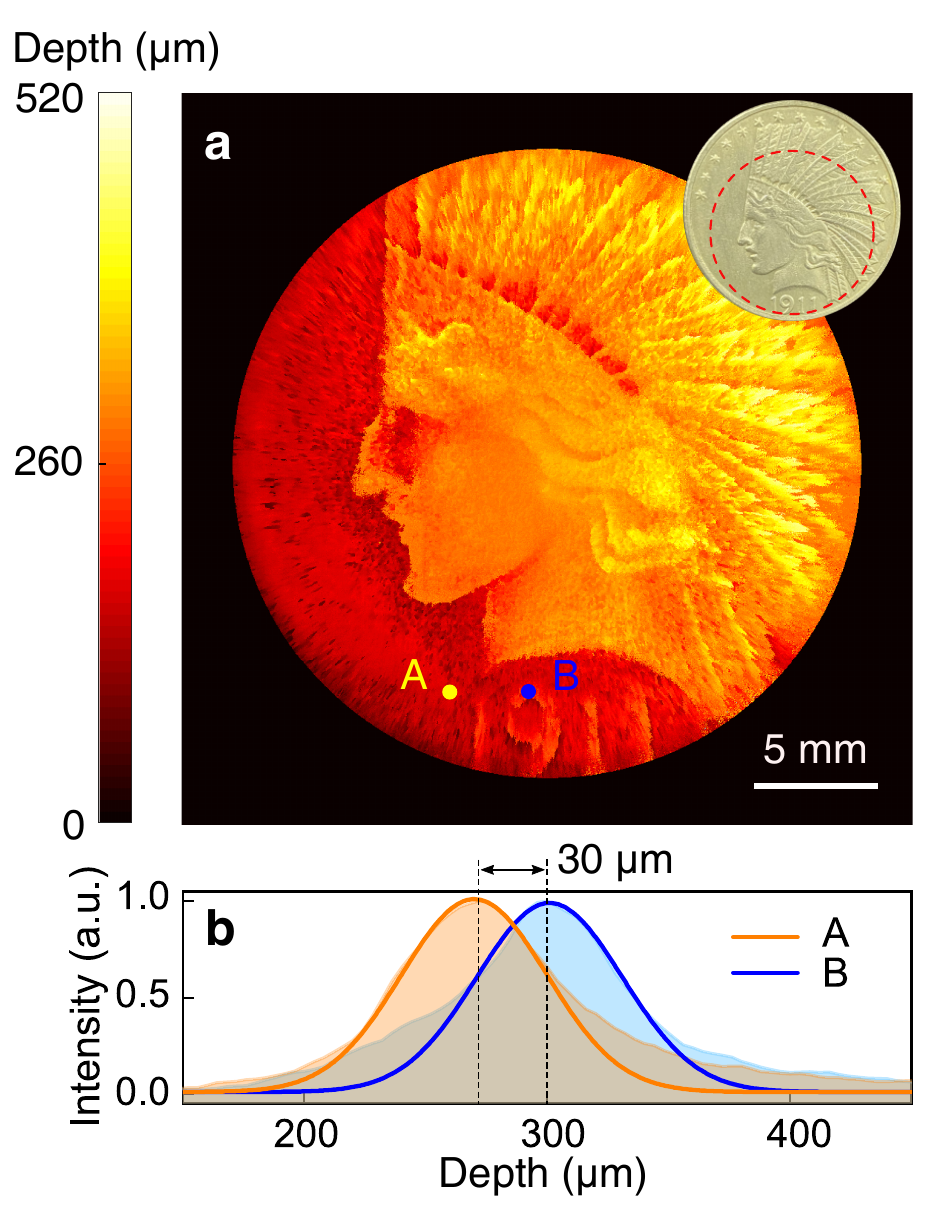}
\caption{\textbf{Tomographic MIR imaging for the structured surface of a coin.} (a) Reconstructed 3D profile for the subtly embossed pattern on the coin surface. Note that the coin is placed under a silicon wafer with a thickness of 5 mm as a visibly opaque obscurant. The ultrafast nonlinear gating facilitates an effective signal screening from the disturbed noises due to high reflections along the penetration path. The photographic image of the coin is presented in the inset, while the red dashed line denotes the imaging area. (b) Cross-correlation traces corresponding to the positions A and B, which indicate a small depth difference of 30 $\mu$m from the peaks of Gaussian fittings in solid lines. Experimental data are illustrated in shaded areas.}
\label{fig2}
\end{figure}

\vspace{8pt}
\noindent\textbf{High-resolution MIR tomographic imaging.} Now we turn to characterize the performance of the implemented MIR 3D imaging system. The depth resolving capability is manifested by profiling a mental coin placed under a silicon wafer with a thickness of 5 mm. One desirable feature for the MIR-based tomography is the reduced light scattering in comparison to the visible counterparts, which thus permits a deep-penetration imaging through dense and thick materials \cite{Potma2021Optica}. The Fresnel reflection of the silicon wafer is about 30\%, which results in a total power transmission of 24\% for four interfaces in the double-pass configuration. Figure \ref{fig2}(a) gives the reconstructed depth map for the coin subtly embossed with a portrait. The depth information for each pixel is inferred from the peak position of the measured cross-correlation trace by scanning the pump delay. Two representative points are denoted with A and B. The intensity correlations are measured at a step of 5 $\mu$m over a scanning depth range of 0.52 mm. As shown in Fig. \ref{fig2}(b), the slight height difference of 30 $\mu$m can clearly be identified for the two targeted locations. The precision of the depth measurement is determined by the accuracy for the peak fitting, which allows to differentiate a height variation at the $\mu$m level. The presented ultrafast optical gating offers a temporal resolution far beyond the sub-nanosecond electronic gate for pixelated detectors, such as SPAD arrays \cite{Gariepy2015NC,Shin2018NC,Morimoto2020Optica} and ICCDs \cite{Morris2014NC,Faccio2020NRP}.

Compared to point-scanning fashion, the wide-field imaging modality significantly reduces the acquisition time of the volumetric data. Here, the exposure time is set to 50 ms for each frame with 1024$\times$1024 pixels. The field of view for the single-shot imaging reaches to a diameter about 2.5 cm. The achieved megapixel definition allows to capture massive elements in parallel, which is particularly useful to realize high-speed sectioning for intricately structured objects. The spatial resolution is estimated to be about 60 $\mu$m based on the imaging performance for a USAF-1951 resolution target (see Supplementary Note 3), which represents at least two-fold improvement over the previous wide-field upconversion imaging performance \cite{Huang2022NC}. The better spatial resolution is partially ascribed to the larger transverse section of 2$\times$3 mm$^2$ for the CPLN crystal, which permits the use of a larger pump size to address more spatial-frequency components at the Fourier plane. Another enhancing factor of $\sqrt{2}$ lies in a shaper convolution kernel for the incoherent imaging paradigm \cite{Barh2019AOP}. Further enhancement of the spatial resolution can resort to a stronger signal focusing within a shorter crystal, which supports a larger acceptance angle for collecting more high spatial-frequency components (see Materials and methods). Meanwhile, the reduced crystal length will also alleviate the temporal walk-off effect, thus allowing to use a shorter pulses to improve the axial resolution. The resulting penalty of a lower conversion efficiency due to the shorter nonlinear interaction length can be compensated by augmenting the pump power. Therefore, the implemented tomographic MIR imaging is featured with wide field, high definition and precise timing, which provides a promising alternation to MIR OCT for the characterization of structured materials.

\begin{figure*}[t!]
\centering
\includegraphics[width=0.95\textwidth]{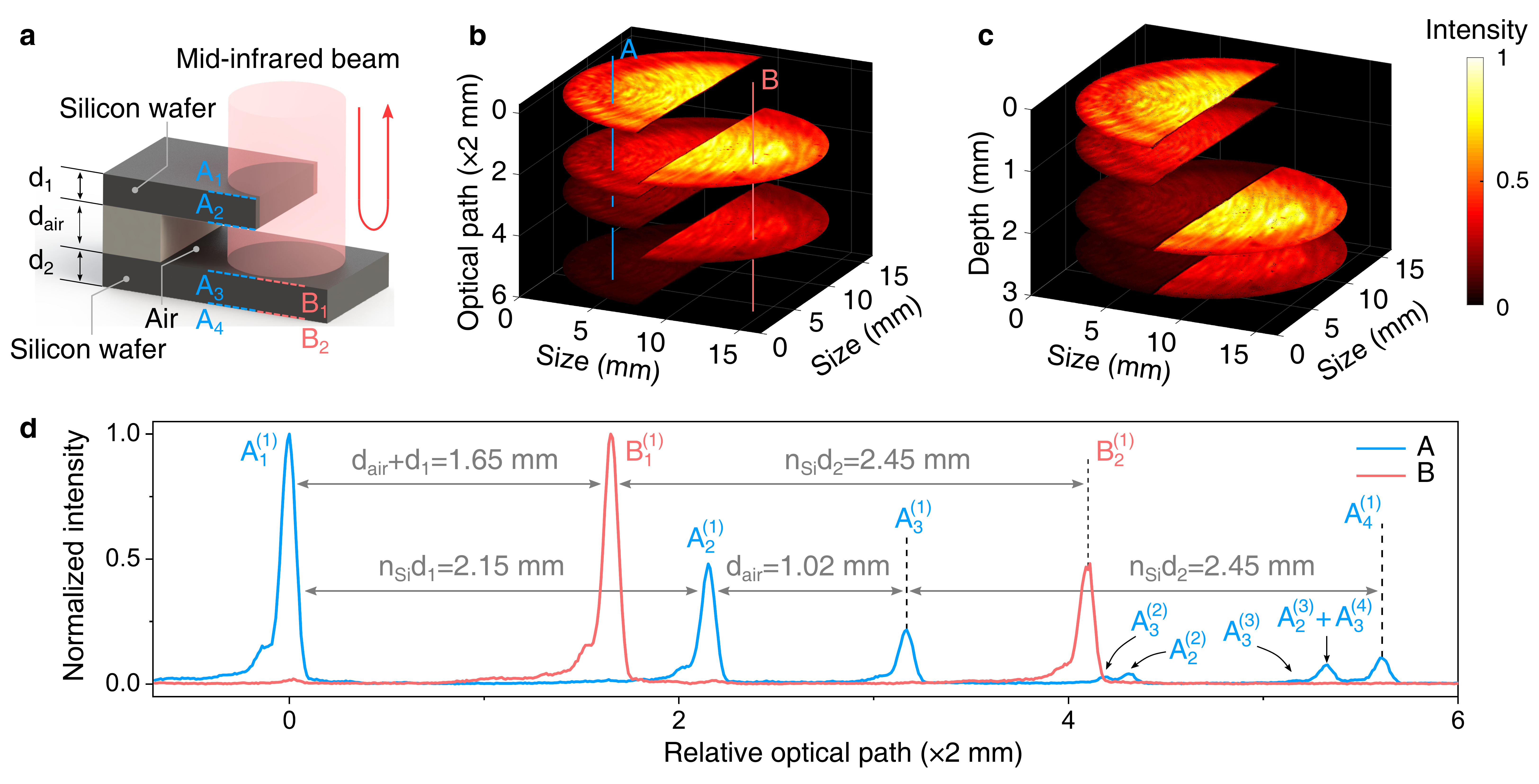}
\caption{\textbf{MIR volumetric imaging for multi-interface structures.} (a) Schematic of a multi-layer sample with two silicon wafers separated by a small air gap. The MIR beam penetrates into the semiconductor material, which allows to interrogate a series of interfaces through the Fresnel reflection. (b) Reconstructed wafer/air interfaces at various optical-path positions relative to the top surface A$_1$. (c) Reconstructed 3D structure after correcting the refractive index for the silicon, which illustrates the geometric architecture of the stacked sheets. (d) Recorded cross-correlation traces at the two positions A and B. Values are given for the relative optical paths between dominant peaks. The positions and amplitudes for these peaks are useful to retrieve the information of the silicon wafer thickness, the material refractive index, and the surface reflectivity. Note that the values of folded optical paths are given due to the reflection imaging configuration, which favors to infer the physical depth by simply correcting the refractive index of the propagating medium.}
\label{fig3}
\end{figure*}

\vspace{8pt}
\noindent\textbf{MIR volumetric imaging for layered structures.} Next, we investigate the volumetric imaging performance for two stacked silicon wafers as a multiple-layer object. The MIR light favors to penetrate through the semiconductor material, such that the back-scattered photons from the interior interfaces can provide rich information on the reflectivity and depth of the imbedded surfaces, as well as the refractive index of the medium. Figure \ref{fig3}(a) illustrates the schematic of the layered sample that is comprised of two silicon wafers with thicknesses of 630 and 720 $\mu$m, respectively. The two wafers are separated by an air gap of 1.02 mm. The MIR beam is bisected by the top wafer, and subsequently reaches the bottom one. The MIR probe is reflected at all the medium/air interfaces due to the refractive-index discontinuity. To capture the multiple reflections, the pulsed optical gate is precisely tuned with an axial step of 10 $\mu$m. The resultant cross-correlation traces for two representative points A and B are presented in Fig. \ref{fig3}(d), which show recorded peaks within a scanning optical path over 12 mm. The depth origin is arbitrarily defined at the surface A$_1$ of the top layer. Each peak is labeled with X$_{m}^{n}$, where X$\in \left\{\text{A, B}\right\}$ and $m$ represent the surface group and surface number as specified in Fig. \ref{fig3}(a), and $n$ indicates the $n^\text{th}$ reflection from the same surface according to an ascending order of the returning time. The relative optical paths between main peaks are denoted with specific numbers, which enable us to identify the positions for the involved surfaces of the stacked wafers.

Consequently, the reconstructed wafer/air interfaces are depicted in Fig. \ref{fig3}(b), which exhibits a dislocation between the two complementary semicircle areas of the bottom wafer. This dislocation is determined by the optical-path difference induced by the top silicon wafer \cite{Mrejen2020LPR}. As a result, the precise knowledge of the axial offset can be used to infer the refractive index of the medium, which permits to reconstruct the geometric structure of the stacked wafers as illustrated in Fig. \ref{fig3}(c). The intensity distribution for each semicircle slice is normalized to the maximum value of the reflection from the surface A$_1$. The dimmer brightness for the subjacent layers is due to the increased overall loss of Fresnel reflections. As shown in Fig. \ref{fig3}(d), several weak peaks occur at longer optical distances in the correlation trace, which correspond to the back-and-forth bounces between the interfaces for the MIR ballistic photons. The detailed traveling paths are presented in Supplementary Note 4.

\begin{figure*}[t!]
\centering
\includegraphics[width=0.95\textwidth]{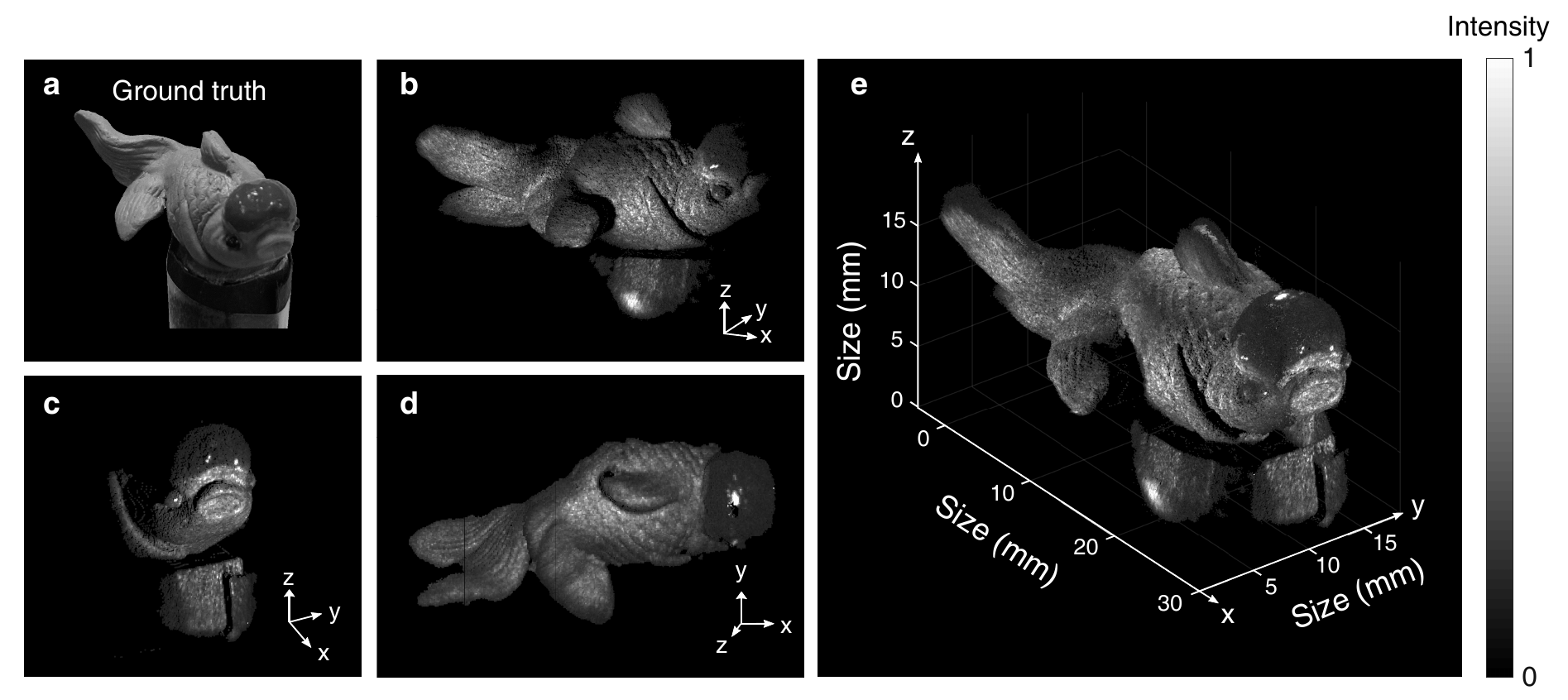}
\caption{\textbf{MIR stereoscopic imaging.} Multiple 3D measurements are conducted from different imaging perspectives, which enable us to fully estimate the 3D coordinates of all points on an object. (a) Ground truth for the targeted object, which is a goldfish made of ceramic. (b-d) Reconstructed 3D images at various illumination directions. Note that a larger field of view is obtained by stitching two adjacent imaging areas, as shown in the side (b) and top (d) views. (e) Stereoscopic illustration of the reconstructed 3D object based on the numerical superimposition of all the estimated surface points in the Cartesian coordinates. Note that the MIR illumination is set at the low-light level such that the average detected photon number per second is about one for each pixel. An stereo-structure overview is given in Supplementary Video 1.}
\label{fig4}
\end{figure*}

In comparision to OCT measurements via the spectral interferometry \cite{Vanselow2020Optica}, the nonlinear optical gating based on the intensity correlation mitigates the stringent requirement of the phase stability for the interacting optical fields. The depth-resolving functionality here is thus more resistant to ambient perturbations such as temperature change and pressure variation of the sample. Another unique feature for the range-gated operation is the possibility to selectively interrogate an particular section within a volumetric target. Moreover, instead of a linear scanning, the use of a flexibly warped trajectory allows to efficiently allocate more longitudinal samples to the area of interest \cite{Jiang2020NP}. Such a foveated operation provides a potential solution to the big data predicament in 3D imaging. Pertinent to the transparency window for semiconductor and polymer materials, the MIR volumetric imaging would open an effective way to perform non-destructive defect inspection for electronic chips and circuit boards \cite{Israelsen2019LSA, Potma2021Optica, Vanselow2020Optica}.

\begin{figure*}[t!]
\centering
\includegraphics[width=1\textwidth]{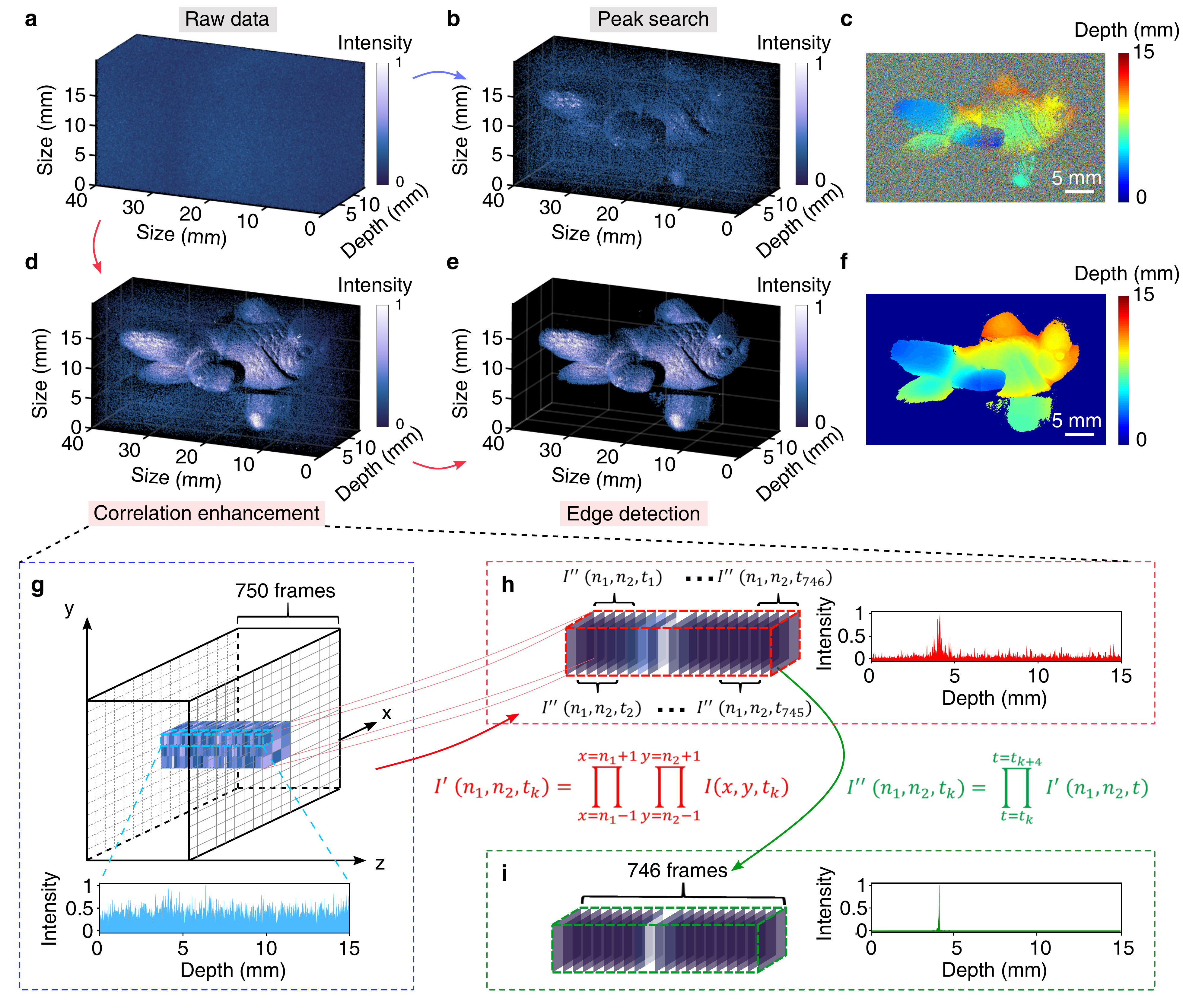}
\caption{\textbf{MIR noise-tolerant 3D imaging in the low-photon-flux regime.} (a) Raw point cloud generated by 750 frames over a depth range of 15 mm. Each frame contains a pixel number of 1642$\times$1024 by stitching two overlapped regions for fully capturing the object profile. The detected photon numbers per second for the signal and noise are about 0.05 and 1000 per pixel in average, respectively. (b) Reconstructed 3D image by selecting the points corresponding to intensity peaks along the depth axis. (c) Depth map obtained by the peak search method. (d) Filtered point cloud based on a correlation-enhancement algorithm for the depth extraction at an extremely low signal-to-noise ratio. (e) Final 3D imaging result after a denoising operation by the edge detection. (f) Depth map of the final 3D reconstruction. (g-i) Detailed algorithmic process for the involved signal extraction based on the correlation enhancement in spatial and temporal dimensions. The continued product of 3$\times$3 elements centered at every pixel in the frame facilitates to reveal the peak position for retrieving the depth information. The contrast of identified cross-correlation peak is further improved by another continued-product operation on the pixel intensities of every five sequential frames. The effectiveness of the denoising operation is verified by the representative correlation traces.} 
\label{fig5}
\end{figure*}

\vspace{8pt}
\noindent\textbf{Single-photon MIR 3D imaging.} In the following, we proceed to investigate the high-sensitivity competence of the MIR 3D imager. The implemented parametric upconversion imaging leverages the silicon-based EMCCD with high spatial resolution and single-photon response, which hence provides high-definition and high-sensitivity performances beyond currently available MIR focal plane arrays \cite{Razeghi2014RPP}. The imaging sensitivity of the EMCCD is optimized at a low thermoelectrically-cooled temperature of -80 $^\circ$C and a high on-chip multiplication gain of 1000. The resultant dark noise is measured to be about 8.7$\times$10$^{-4}$  electrons/pixel/second. It is the megapixel and sensitive imaging capability that makes it possible to conduct a wide-field MIR tomographic analysis at the photon-starving regime. As shown in Fig. \ref{fig4}(a), the test object is a ceramic goldfish with both geometric and reflectivity complexity, which is placed on the top of an optical post. The post is wrapped by a piece of black paper to avoid the mirror reflection due to the shining mental surface. The MIR illumination is diminished to the low-light level such that the detected photon number per second is about one for each pixel in average.

In the experiment, the 3D depth imaging is performed at different perspectives in order to illuminate the whole surface of the target. The reconstructed grey-scale images at side, front and top projections are presented in Figs. \ref{fig4}(b-d). Each set of cube data contains 750 frames with a depth step of 20 $\mu$m, and the exposure time for each frame is set to 2 s. The collective information from the multiple 3D measurements allows us to map all the sampled points on the object into the 3D space. Specifically, the reconstructed surfaces from various perspectives are stitched together to realize the MIR stereoscopic imaging at the single-photon level, as illustrated in Fig. \ref{fig4}(e). Thanks to the high depth resolution, a great deal of object details can be recovered, such as fish scale, as well as line textures on the fins and tail. Comparing to conventional techniques based on shape of shading or photometric stereo \cite{Sun2013Science, Martin2013NM}, the presented stereoscopic approach eliminates the requirement of sophisticated algorithms for the imaging reconstruction. The all-round visualization of the 3D scene is recorded in Supplementary Video 1. The presented stereoscopic vision favors to examine 3D samples at an arbitrary observation angle, which may facilitate the target recognition and identification \cite{Martin2013NM}.

Finally, we have further investigated the noise-tolerant imaging ability in the condition of ultra-low SNRs. To this end, the MIR probe light is further attenuated such that the average detected flux is only 0.05 photons/second/pixel, which is more than four orders of magnitude smaller than the background noise of 1000 photons/second/pixel. The signal and noise contributions are derived from the volumetric photon-counting dataset with dimensions of 1024$\times$1024$\times$750. The integration time for each frame is set to be 10 s for collecting sufficient photons. Figure \ref{fig5}(a) presents point cloud for the acquired raw data, where the signal is indeed overwhelmed by the severe noise. In this scenario, the simple peak-search processing barely reveals the object shape, as shown in the resulting intensity distribution in Fig. \ref{fig5}(b). The inability to accurately identify the depth positions is also manifested by the random noises in the depth map in Fig. \ref{fig5}(c). The discontinued depth distribution with sharp and sudden disturbances is ascribed to ill-identified peak positions of the noisy correlation traces.

To better recover the object shape and surface reflectivity, a numerical denoiser is implemented by exploiting the structural information of the target in both the transverse and longitudinal domains to censor extraneous contributions from background noise and dark counts. The underlying procedures for the spatio-temporal correlation enhancement are elaborated in Figs. \ref{fig5}(g-i). Each pixel value in the frame is replaced with the continued product of the 3$\times$3 region consisting of the pixel and its eight neighboring elements. As a result, the peak profile of the intensity correlation becomes visible as shown in Fig. \ref{fig5}(h). The contrast of the intensity-correlation trace is subsequently enhanced in the axial dimension by performing another continued-product operation on the pixel intensities of every five sequential frames. Consequently, the depth information at the chosen pixel can be precisely retrieved from the high-contrast cross-correlation peak as shown in Fig. \ref{fig5}(i). The effectiveness of the denoising approach is verified by the filtered point cloud in Fig. \ref{fig5}(d). Figure \ref{fig5}(e) presents the final 3D imaging result after cleaning the residual noises by the edge detection, while the corresponding depth map is shown in Fig. \ref{fig5}(f). More details on the involved algorithm and numerical implementation are given in Supplementary Note 5. 

In these measurements, the depth range of interest is set to be 15 mm, corresponding to a temporal window of 100 ps. The implemented ultrashort analysis precision enabled by the femtosecond optical gating is far beyond the timing resolution about tens of picoseconds for other single-photon imagers based on Geiger-mode avalanche photodiodes \cite{Gariepy2015NC, Shin2018NC} or superconducting nanowire sensors \cite{Kong2020OL}. Moreover, the range-gated operation facilitates time-resolved measurements without being impeded by the pileup distortions, dead time, and count-rate saturation issues \cite{Rehain2020NC}, which contrasts with the tight-of-flight 3D imaging technique based on time-correlated single-photon counting (TCSPC) \cite{Gariepy2015NC, Shin2018NC, Kong2020OL}. The presented low-light MIR 3D imager with a high timing resolution would facilitate important photon-starving applications, such as biological imaging and covert imaging, where a high photon flux would have detrimental effects \cite{Morris2014NC}.

\vspace{8pt}
\noindent{\fontfamily{phv}\selectfont 
\textbf{Discussion}}
\vspace{4pt}
\newline
We have implemented a high-performance 3D imaging system at the MIR spectral region, which is featured with single-photon sensitivity, femtosecond timing resolution, and wide-field operation. Particularly, the combination with a denoising algorithm based on spatiotemporal correlation enables us to perform a high-contrast recovery of 3D scenes under an extremely low SNR of 5$\times$10$^{-5}$, which addresses the MIR imaging challenge in both photon-starved and noise-polluted regimes. The single-photon sensitive 3D imager is highly demanded in various low-light scenarios, for instance phototoxicity-free examination of biological specimens, non-destructive characterization of photosensitive chemical materials, or remote sensing of a dynamic scene at a long standoff distance \cite{Altmann2018Science}. 

Moreover, the superior temporal resolution along with the high detection sensitivity renders our imaging system attractive for time-gated ballistic-photon MIR tomography. In this case, the ultrashort time gate is useful to suppress multiply scattered photons, which can significantly improve the visualization of objects hidden in the highly scattering medium \cite{Wang1991Science} or beyond the line of sight \cite{Faccio2020NRP}. Notably, the presented 3D imaging modality is ready to include the spectral dimension based on a tunable MIR light source \cite{Potma2021Optica, Junaid2019Optica}. The envisioned MIR spectro-tomography is foreseen to provide label-free, non-destructive 3D visualizations of biological and materials samples.

\vspace{8pt}
\noindent  {\fontfamily{phv}\selectfont 
\normalsize \textbf{Materials and methods}}

\noindent \textbf{\textsf{Details on imaging setup.}} The YDFL at 1030 nm is a mode-locked fiber laser, which delivers ultrafast pulses at the repetition rate about 21.6 MHz. The output power is boosted to serve the pump source. The pulse duration is measured to be 270 fs with an optical auto-correlator. The ECDL operates at the continuous-wave mode, and can be spectrally tuned from 1527 to 1565 nm. A MIR signal source is prepared with a PPLN crystal through DFG. The pulse duration of the MIR pulse is inferred to be 295 fs based on cross-correlation technique. The MIR source is self-synchronized with the pump, which facilitates the coincident-gating upconversion imaging within a CPLN crystal. The CPLN is fabricated with geometric dimensions of 2$\times$3$\times$10 mm$^3$ in terms of thickness$\times$width$\times$length. The poling period linearly ramps from 16 to 24 $\mu$m along the axial direction, which is designed to support a wide-field parametric upconversion imaging over 3-5 $\mu$m. To remove the pump-induced fluorescence noise, the upconverted signal at 771 nm passes through a spectral filtering group with a total power transmission of 80\% and a noise rejection ratio up to 210 dB. The filtered SFG image is finally captured by an EMCCD (Andor, iXon Ultra 888). The silicon-based camera is specified with a detection efficiency at 771 nm of 80\%, a spatial resolution of 13 $\mu$m, and a pixel number of 1024$\times$1024. More information about experimental setup is given in Supplementary Note 1. \\

\noindent \textbf{\textsf{Characterization of imaging performance.}} In the 4-f upconversion imaging system, the spatial resolution is determined by the aperture size of the crystal or the beam size of the pump whichever is the smaller. In the experiment, the beam diameter of the pump approaches to the transverse section of the crystal, which allows to obtain an optimized spatial resolution about 60 $\mu$m in the incoherent imaging regime. The resolution can be further enhanced by using a shaper focus of the signal beam or a larger section of the nonlinear crystal. The longitudinal resolution of the 3D imaging system is defined by the peak width of the cross-correlation trace between the signal and pump pulses. The peak width is measured to be 400 fs, corresponding to a depth resolution of 60 $\mu$m. The involved time scanning is realized by a translational stage (Thorlabs, NRT150/M) with an on-axis accuracy of 2 $\mu$m. Notably, a much smaller depth variation can be identified by fitting the measured intensity-correlation traces, which may differentiate a axial change at the $\mu$m level. See Supplementary Notes 2 and 3 for more discussions.\\

\noindent \textbf{\textsf{Algorithms for imaging reconstruction.}} The 3D imaging is performed in a wide-field fashion, thus eliminating the need of a time-consuming raster scan. Each frame is acquired with 1024$\times$1024 points. Depending on the illumination intensity, the exposure time of the camera varies from 10 $\mu$s to 10 s. In contrast to direct imagers, the temporal resolution is not limited by the device, but determined by the ultrafast optical gating. The time stamp for each frame is registered as the pump delay. Consequently, the knowledge of the 3D coordinate for each point allows us to retrieve the information of structure and reflectivity. The surface is revealed at depth positions with maximum intensities along the axial direction. Notably, the optical slicing capability facilitates flexible observation on an arbitrary section of the volumetric scene. In the low-photon-flux regime, a numerical denoiser is implemented to reveal the signal under severe noises, which exploits  the strong correlation in the spatial and time domains for a natural scene. The involved algorithm includes mainly two steps of correlation enhancement and edge detection. The numerical computation is performed by using MATLAB in a laptop with an Intel i5-8300H processor running at 2.3 GHz. Detailed descriptions about the data processing are presented in Supplementary Note 5.

\vspace{8pt}
\noindent  {\fontfamily{phv}\selectfont 
\normalsize \textbf{Acknowledgements} 
}
\newline
\noindent This work was supported by National Natural Science Foundation of China (62175064, 62235019, 62035005); Shanghai Pilot Program for Basic Research (TQ20220104); Shanghai Municipal Science and Technology Major Project (2019SHZDZX01); Fundamental Research Funds for the Central Universities.

\vspace{8pt}
\noindent  {\fontfamily{phv}\selectfont 
\normalsize \textbf{Author contributions} 
}
\newline
\noindent K.H., J.F., and H.Z. conceived the project and designed the experiments. J.F. and K.H. built the system, performed experiments, and processed data. M.Y. built fiber laser sources. E W. analyzed the imaging data. J.F. and K.H. wrote the manuscript draft. All authors were involved in discussions and contributed to the manuscript editing.

\vspace{8pt}
\noindent  {\fontfamily{phv}\selectfont 
\normalsize \textbf{Data availability} 
}
\newline
\noindent The data that support the findings of this study are available from the corresponding author upon reasonable request.

\vspace{8pt}
\noindent  {\fontfamily{phv}\selectfont 
\normalsize \textbf{Conflict of interest} 
}
\newline
\noindent The authors declare no competing interests.

\vspace{8pt}
\noindent  {\fontfamily{phv}\selectfont 
\normalsize \textbf{Supplementary information} 
}
\newline
\noindent The online version contains supplementary material available at https://doi.org/XXXXX.

\end{document}